\documentclass[pre,preprint,showpacs,floatfix,superscriptaddress]{revtex4}
\usepackage{graphicx,amsmath,amsfonts,float}
\usepackage{undertilde}
\newcommand*{\et}[0]{\textit{et~al.}}
\newcommand*{\G}[0]{\protect\utilde{G}}
\newcommand*{\GR}[0]{\protect\underline{G}}

\newcommand*{\D}[0]{\underline{D}}
\newcommand*{\tD}[0]{\protect\utilde{D}}
\newcommand*{\Gii}[0]{\protect\utilde{G}^{II}}
\newcommand*{\Gib}[0]{\protect\utilde{G}^{I\sigma}}
\newcommand*{\Dinv}[0]{(\protect\utilde{D}^{II}(\kx))^{-1}}
\newcommand*{\R}[0]{\vec{R}}
\newcommand*{\Rp}[0]{\vec{R}^{\prime}}
\newcommand*{\zp}[0]{z^{\prime}}
\newcommand*{\xp}[0]{x^{\prime}}
\newcommand*{\kv}[0]{\vec{k}}
\newcommand*{\khat}[0]{\hat{k}}
\newcommand*{\kx}[0]{k_{x}}

\newcommand*{\kz}[0]{k_{z}}
\newcommand*{\xl}[0]{x_{l}}
\newcommand*{\xlp}[0]{x_{l'}}
\newcommand*{\zl}[0]{z_{l}}
\newcommand*{\zlp}[0]{z_{l'}}
\newcommand*{\f}[0]{\vec{f}}
\newcommand*{\uv}[0]{\vec{u}}
\newcommand*{\tone}[0]{\vec{t}_{1}}
\newcommand*{\ttwo}[0]{\vec{t}_{2}}
\newcommand*{\eqn}[1]{Eqn.~(\ref{eqn:#1})}
\newcommand*{\fig}[1]{Figure~\ref{fig:#1}}
\newcommand*{\tab}[1]{Table~\ref{tab:#1}}

\begin{document}

\title{Lattice Green's function for crystals containing a planar interface}
\author{M. Ghazisaeidi}
\affiliation{Department of Mechanical Science and Engineering, University of Illinois at Urbana-Champaign, Urbana, Illinois 61801, USA}
\author{D. R. Trinkle}
\affiliation{Department of Materials Science and Engineering, University of Illinois at Urbana-Champaign, Urbana, Illinois 61801, USA}

\begin{abstract}
    Flexible boundary condition methods couple an isolated defect to a harmonically responding medium through the bulk lattice Green's function; in the case of an interface, interfacial lattice Green's functions. We present a method to compute the lattice Green's function for a planar interface with arbitrary atomic interactions suited for the study of line defect/interface interactions. The interface is coupled to two different semi-infinite bulk regions, and the Green's function for interface-interface, bulk-interface and bulk-bulk interactions are computed individually. The elastic bicrystal Green's function and the bulk lattice Green's function give the interaction between bulk regions. We make use of partial Fourier transforms  to treat in-plane periodicity. Direct inversion of the force constant matrix in the partial Fourier space provides the interface terms. The general method makes no assumptions about the atomic interactions or crystal orientations. We simulate a screw dislocation interacting with a $(10\bar{1}2)$ twin boundary in Ti using flexible boundary conditions and compare with traditional fixed boundary conditions results. Flexible boundary conditions give the correct core structure with significantly less atoms required to relax by energy minimization. This highlights the applicability of flexible boundary conditions methods to modeling defect/interface interactions by \textit{ab initio} methods. 
\end{abstract}

\maketitle

\section{Introduction}
Accurate atomic scale studies of lattice defect geometry is the key to any
modeling of their effects on material properties.  However, the long-range
(elastic) displacement field of isolated defects, e.g., dislocations, is
incompatible with periodic boundary conditions typically used in computer
atomistic simulations.  Fixed boundary conditions require simulation sizes
large enough for the elastic solution to be accurate---a size typically
beyond even modern density-functional theory methods.  
Flexible boundary condition methods avoid these issues by relaxing the
atoms away from the defect core through lattice Green's function (LGF) as
if they are embedded in an infinite harmonic medium.  Hence, the atomic scale
geometry of the defect core is coupled to the long-range strain field in
the surrounding medium.  Sinclair \et\ introduced flexible boundary
conditions for studying defects in bulk materials\cite{ref:Sinclair} such
as cracks\cite{Thomson:rel,crack},
dislocations\cite{Rao-dislocation,Rao-screw,Yang-screw,Rao-slip},
vacancies with classical potentials and isolated screw or edge dislocations
with density-functional
theory\cite{Rao-screw-DFT,ref:woodward,ww-screw-DFT,al-prl}. Flexible boundary
conditions use the LGF corresponding to the specific geometry of the
problem.  For instance, line defects in the presence of interfaces require
the interfacial lattice Green's function (ILGF).  Line defects in
interfaces affect the mechanical properties of composites, two-phase or
polycrystalline materials where heterophase or homophase interfaces
interact with defects.  Tewary and Thomson\cite{ref:tewary_lgf} proposed a
Dyson-equation calculation of the interfacial lattice Green's function
suitable for materials with short-range atomic interactions and simple
crystal structures.  We present a general---for all types of interactions
and interface orientations---accurate method to compute the interfacial
lattice Green's function, suited to use in density functional theory.
Specifically, this method is applicable to studies of line defects
interactions with planar interfaces such as disconnections in interfaces
and dislocation or crack tips interacting with grain boundaries and
two-phase interfaces.  We compute the Green's function
for a $\left(10\bar{1}2\right)$ twin boundary in Ti to simulate a screw
dislocation interacting with the twin boundary using flexible boundary
conditions.  Section~\ref{sec:harmonic_response} reviews the harmonic
response functions: the force constant matrix and the lattice Green's
function.  Section~\ref{sec:interfacialGF} explains the general procedure
for evaluation of the interfacial lattice Green's function and
section~\ref{sec:application} applies the method to modeling the
interaction of a screw dislocation with Ti $\left(10\bar{1}2\right)$ twin
boundary.  The end result is a computationally tractable, general approach
usable for studies of defects in interfaces.

\section{Harmonic Response}
\label{sec:harmonic_response}
Harmonic response is characterized by a linear relationship between forces
and displacements\cite{ref:harmonic}. Lattice Green's function $\GR(\R,\Rp)$ relates the
displacement $\uv(\R)$ of atom $\R$ to the internal forces $\f(\Rp)$ on
another atom $\Rp$ of the crystal through
\begin{equation}
       \uv(\R)=-\sum_{\Rp}\GR(\R,\Rp)\f(\Rp).
\label{eqn:GFdef}
\end{equation}
Conversely, the forces on an atom can be expressed in terms of
displacements through the force constant matrix $\D(\R,\Rp)$ by
\begin{equation}
   \f(\R)=-\sum_{\Rp} \D(\R,\Rp) \uv(\Rp).
\label{eqn:Ddef}
\end{equation}
Translational invariance of an infinite crystal makes $\GR$ and $\D$
functions of the relative positions of the atoms.  Substituting
\eqn{Ddef} into \eqn{GFdef} gives
$\displaystyle\sum_{\Rp}\GR(\R-\Rp)\D(\Rp)=\textbf{1}\delta(\R)$, where
$\delta(\R)$ is the Kronecker delta function.  A constant shift in atom
positions does not produce internal forces; hence, $\sum_{\R}{\D(\R)}=0$,
and so $\GR(\R)$ is the pseudo inverse of $\D(\R)$ in the subspace without
uniform displacements or forces.  In a bulk geometry, the Fourier transform
of the lattice functions are defined as
\begin{equation*}
  \G(\kv)=\sum_{\R}e^{i\kv\cdot\R}\GR(\R),\quad \GR(\R)=\int_{BZ}\frac{d^{3}k}{(2\pi)^{3}}e^{-i\kv\cdot\R}\G(\kv)
\end{equation*}
where the summation is over lattice points.  In reciprocal space, the
matrix inverse relation $\G(\kv)\tD(\kv)=1$ and the sum rule
$\tD(\vec{0})=0$ require that $\G(\kv)$ has a pole at the
$\Gamma$-point. While computation of the force constant matrix
$\D(\R)$---and subsequently $\tD(\kv)$---is straightforward, $\GR(\R)$ can
not be computed directly due to its long range behavior. Instead, we
invert $\tD(\kv)$ to get $\G(\kv)$ and then perform
an inverse Fourier transform. Convergence of the inverse Fourier transform
requires an analytical treatment of the pole at the
$\Gamma$-point\cite{ref:lgf,ref:effic}.  In an interface
geometry, translational invariance is broken in the direction perpendicular
to the interface; we use  Fourier transforms in the interface plane
only.  This produces an infinite dimensional dynamical matrix that can not
be simply inverted, but requires a more complex computational approach.

\section{Computation of lattice Green's function for a planar interface}
\label{sec:interfacialGF}

\begin{figure} 
 \begin{center}
  \includegraphics[angle=270,width=6in]{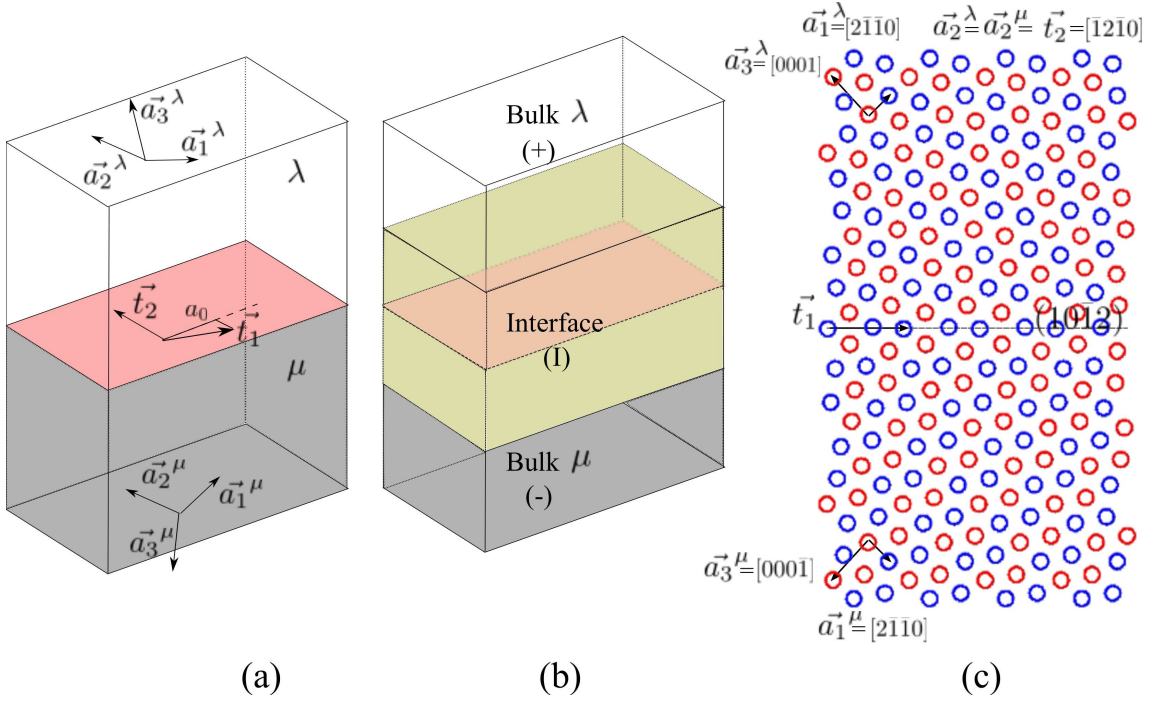}
 \end{center}

\caption{(a) Bicrystal $\mu$ and $\lambda$, (b) separation into bulk and
interface regions and (c) the Ti $\left(10\bar{1}2\right)$ twin boundary .
Two different lattices, $\lambda$ and $\mu$ are connected through a planar
interface. The unit cells of $\lambda$ and $\mu$ are given by
$\vec{a_1}^{\lambda,\mu}$,$\vec{a_2}^{\lambda,\mu}$ and
$\vec{a_3}^{\lambda,\mu}$---all of which must be lattice vectors in
$\lambda$ and $\mu$. The combined lattice has the periodicity of the
interface in $\tone$ and $\ttwo$ directions. Introducing a line defect
threading direction $\ttwo$ reduces the problem to 2D in the plane normal
to $\ttwo$. In (b), the crystal is divided into two semi-infinite bulk
regions, bulk $\lambda$ and bulk $\mu$ symbolized by $(+)$ and $(-)$
respectively, coupled with an interface region (I). The bulk regions are
far from and affected only through an elastic effect by the interface. The
force constant matrix between atom pairs in the bulk is not affected by the
interface. The remaining layers are included in (I). (c) shows the
periodicity vectors for the Ti $\left(10\bar{1}2\right)$ twin boundary. The
interface is defined by $\tone=\sqrt{3a^2+c^2}\hat{x}$ and $\ttwo=a\hat{y}$
where $a$ and $c$ are the hcp unit cell parameters in Ti for both $\lambda$
and $\mu$. $\mu$ is the reflection of $\lambda$ about the interface plane.
}
\label{fig:interface}
\end{figure}

\fig{interface}a shows two lattices, $\lambda$ and $\mu$ joined at a planar
interface. Each set of vectors $\vec{a_1}^{\lambda,\mu}$,
$\vec{a_2}^{\lambda,\mu}$ and $\vec{a_3}^{\lambda,\mu}$ give the periodic
directions in their corresponding lattice.  We introduce integer matrices
$\underline{M}^{\lambda}$ and $\underline{M}^{\mu}$ and deformation
operators $\underline{F}^\lambda$ and $\underline{F}^\mu$ so that
\begin{equation}
\label{eqn:integerm}
 \underline{F}^{\lambda,\mu}
\left[\vec{a_1}^{\lambda,\mu},
      \vec{a_2}^{\lambda,\mu},
      \vec{a_3}^{\lambda,\mu}\right]
\underline{M}^{\lambda,\mu}
=\left[\vec{T_1}^{\lambda,\mu},
       \vec{T_2}^{\lambda,\mu},
       \vec{T_3}^{\lambda,\mu}\right]
\end{equation}
to define the supercell.  We use
$\vec{T_1}^{\lambda}=\vec{T_1}^{\mu}=\tone$ and
$\vec{T_2}^{\lambda}=\vec{T_2}^{\mu}=\ttwo $ as nonparallel vectors to
define the interface plane where $\ttwo$ will be the periodic threading
vector for a line defect in the interface.  The combined lattice has
translational invariance in $\tone$ and $\ttwo$ directions in the interface
plane while the periodicity is broken in directions outside the
plane.  Introducing a threading direction reduces the problem to 2D (i.e
plain strain or anti-plane strain conditions). We confine our calculations
to the plane orthogonal to $\ttwo$ and define the Cartesian coordinate
$\hat{x}$, $\hat{y}$, $\hat{z}$ so that $\tone\cdot\hat{x}=a_0$,
$\ttwo=\left|\ttwo\right|\hat{y}$ and $\hat{z}=\hat{x}\times\hat{y}$. Note
that in general $a_{0}\neq\left|\tone\right|$ because $\tone$ and $\ttwo$
can be nonorthogonal.  Specifically, the lattice positions,
$\R=x\hat{x}+z\hat{z}$ and the Fourier vectors,
$\kv=\kx\hat{x}+\kz\hat{z}$, will be 2D vectors through out this paper and
\begin{equation*}
\D_{\alpha\alpha'}(\R,\Rp)=\D_{\alpha\alpha'}(x-\xp;z,\zp)
\end{equation*}
with $\alpha$ and $\alpha'$ identifying the $xyz$
components
of the second rank tensor $\D$ in Cartesian coordinates.
We index atoms in our computational cell with integer $l$ at position
$(\xl, \zl)$; due to periodicity in the $\hat x$ direction, each atom
also occurs  at $\xl+na_0\hat x$ for integer values of $n$.  The partial
Fourier transform is
\begin{eqnarray}
\label{eqn:pft}
\nonumber
\tD_{\alpha l,\alpha' l'}(\kx) & = & \sum_{n=-\infty}^{\infty}e^{i\kx(\xl-\xlp+na_0)}\D_{\alpha\alpha'}(\xl-\xlp+na_0;\zl,\zlp)\\
\label{eqn:ipft}
\D_{\alpha\alpha'}(\xl-\xlp+na_0;\zl,\zlp)&=&\frac{a_0}{2\pi}\int_{-\pi/a_0}^{\pi/a_0}e^{-i\kx
(\xl-\xlp+na_0)}\tD_{\alpha l,\alpha' l'}(\kx)d\kx
\end{eqnarray}
for all pairs $l,l'$.
Note that ``$l$" indexes layers of atoms with particular z values. There
may be two different layers that have equal $z$: $z_l=z_{l'}$ while
$l\neq l'$. $\tD(\kx)$ is infinite dimensional due to infinite
values of $l$.

To avoid the inversion of infinite dimensional $\tD(\kx)$, the geometry is
divided into two semi-infinite bulk regions coupled with an interface
region. \fig{interface}b shows the schematic divisions of the regions in an
interface geometry consisting of lattices $\lambda$ and $\mu$. The ``bulk"
regions represent layers of atoms that are far from and affected only
through an elastic field by the interface. The atomic scale interaction
between atom pairs are as if they were in their corresponding bulk
geometry. Bulk $\lambda$ and bulk $\mu$ are symbolized by $(+)$ and $(-)$
in our notation. The remaining layers, affected by the reconstructions near
the interface, are included in the ``interface" region (I).  We define the
interface region as atoms where the force constant matrix differ from those
in the bulk lattice.  For specific geometries, additional bulk layers may
be included in the interface to insure a smooth transition between the
regions.  We block partition the infinite dimensional $\tD_{\alpha
l,\alpha'l'}(\kx)$ and $\G_{\alpha l,\alpha'l'}(\kx)$ based on the atom
region (+, $-$, or I) of indices $l$ as
\begin{equation}
\label{eqn:DG}
  \tD(\kx)=
\left(\begin{array}{c|cc}
\tD^{II}(\kx)& \tD^{I-}(\kx)&\tD^{I+}(\kx)\\
\hline
\tD^{-I}(\kx)& \tD^{--}(\kx)&\tD^{-+}(\kx)\\
\tD^{+I}(\kx)& \tD^{+-}(\kx)&\tD^{++}(\kx)
\end{array}\right)
\end{equation}
where $l>l_+$ belong to $(+)$ region, $l<l_-$ belong to $(-)$ region and
the finite-dimensional region is (I). $\tD(\kx)$ and $\G(\kx)$ are
Hermitian and satisfy
\begin{equation}
\label{eqn:identity}
\sum_{\alpha'' l''}
\tD_{\alpha l,\alpha''l''}(\kx)
\G_{\alpha''l'',\alpha'l'}(\kx)
=\delta_{\alpha\alpha'}\delta_{ll'}.
\end{equation}
We construct $\tD(\kx)$ by direct calculation of
$\D_{\alpha\alpha'}(x_l-x_l'+na_0;z_l,z_l')$ followed by a partial Fourier
transform according to \eqn{pft} and block partitioning as in
\eqn{DG}.  Note that due to the finite number of interface layers
and decay of the force constant matrix, the infinite
dimensional non-zero sections of $\tD(\kx)$ consists of $--$, $-+$ and $++$
interactions (bulk-like regions with themselves) which we explicitly avoid
in our approach.

The infinite dimensional blocks of $\G(\kx)$ are known from bicrystal
elastic and bulk lattice calculations. The distance between $+$ and $-$ is
large enough for the elastic Green's function to be applicable; the real
space solution of $\GR^{-+}$ is calculated from the bicrystal elastic
Green's function in both plane strain and anti-plain conditions proposed by
Tewary et.~al\cite{ref:tewary}.  We partially Fourier transform the real
space solution by a continuum version of \eqn{pft},
\begin{equation}
\G^{-+}_{\alpha l,\alpha'l'}(\kx)
=\int_{-\infty}^{\infty}\underline{G}^{-+}_{\alpha\alpha'}(x;z_l,z_{l'})e^{i\kx x}dx.
\label{eqn:gft}
\end{equation}
$\G^{+-}$ is the conjugate transpose of $\G^{-+}$ due to $\G(\kx)$ being
Hermitian.  The functional form of $\GR^{-+}(x;\zl,\zlp)$ consists of real
parts of $\ln(x+p^{\lambda}_{q}\zl+p^{\mu}_{q^\prime}\zlp)$ where $p^{\lambda}_{q}$ and $p^\mu_{q^\prime}$ are the
complex roots of the sextic equation of anisotropic elasticity for bicrystal $\lambda\mu$ and $q,q^\prime=1,2$ in  plain strain and 1 in anti-plane conditions
\cite{ref:tewary}.  We rewrite
$\ln(x+\gamma^{qq^\prime}_{l,l^\prime}+i\beta^{qq^\prime}_{l,l^\prime})$ with 
$$ \gamma^{qq^\prime}_{l,l^\prime}=\Re{(p^{\lambda}_{q})}\zl+\Re{(p^\mu_{q^\prime})}\zlp ,\  \beta^{qq^\prime}_{l,l^\prime}=\Im{(p^{\lambda}_{q})}\zl+\Im{(p^\mu_{q^\prime})}\zlp .  $$
The Green's function in real space is the real part of
the complex logarithm with the form
\begin{equation}
\GR^{-+}_{\alpha\alpha^\prime}(x;\zl,\zlp)=\sum_{q,q^\prime} \underline{a}_{\alpha\alpha^\prime}^{qq^\prime}\ln\left|(x+\gamma^{qq^\prime}_{l,l^\prime})^2+(\beta^{qq^\prime}_{l,l^\prime})^2\right|+\underline{b}_{\alpha\alpha^\prime}^{qq^\prime}\arctan\left(\frac{\beta^{qq^\prime}_{l,l^\prime}}{x+\gamma^{qq^\prime}_{l,l^\prime}}\right)
\label{eqn:log}
\end{equation}
where  $\underline{a}_{\alpha\alpha^\prime}^{qq^\prime}$ and $\underline{b}_{\alpha\alpha^\prime}^{qq^\prime}$ are real valued coefficients of the term $qq^\prime.$
\eqn{log} is obtained by rewriting Eqn. (60) in \cite{ref:tewary}.
The partial Fourier transform is
\begin{equation}
\G_{\alpha l,\alpha'l'}^{-+}(\kx)
=-\frac{\pi}{\left|\ttwo\right|\left|\kx\right|}\sum_{q,q^\prime} 2\underline{a}_{\alpha\alpha^\prime}^{qq^\prime}e^{-\left|\beta^{qq^\prime}_{l,l^\prime}\kx\right|}e^{-i\gamma^{qq^\prime}_{l,l^\prime}\kx}+i \underline{b}_{\alpha\alpha^\prime}^{qq^\prime}\frac{\kx}{\left|\kx\right|}e^{-\left|\beta^{qq^\prime}_{l,l^\prime}\kx\right|}e^{-i\gamma^{qq^\prime}_{l,l^\prime}\kx}
\label{eqn:bipft}
\end{equation}
with a first order pole at $\kx=0$. The $1/\left|\ttwo\right|$ prefactor is required for the elastic and lattice Green's functions to have consistent units of ($\text{length}^2/\text{energy}$).  
We separate the pole from the remainder of the Green's function
\begin{equation}
\label{eqn:elas_sep}
\G^{-+}_{\alpha l,\alpha'l'}(\kx)=
\frac{\hat{\G}_{\alpha\alpha'}^{-+}}{\left|\kx\right|}
+\check{\G}_{\alpha l,\alpha'l'}^{-+}(\kx).
\end{equation}
The pole with a constant coefficient $\hat{\G}^{-+}_{\alpha\alpha'}=-\frac{\pi}{\left|\ttwo\right|}\sum_{q,q^\prime} \underline{a}_{\alpha\alpha^\prime}^{qq^\prime}$ will
be treated analytically while the nonsingular remainder $\check{\G}^{-+}_{\alpha
l,\alpha'l'}(\kx)$, will be treated numerically.

The $\G^{--}(\kx)$ and $\G^{++}(\kx)$ blocks in \eqn{DG} are obtained from the bulk
lattice Green's function of $\lambda$ and $\mu$ lattices plus an elastic
term due to the presence of the interface.  The full Fourier transform of
the bulk LGF $\G^{\sigma\sigma}(\kv)$ is the inverse of the bulk dynamical
matrix from Section~\ref{sec:harmonic_response}.
The partial inverse Fourier transform gives the Green's function in
terms of $\kx$ and atom indices
\begin{equation}
\G_{\alpha l,\alpha'l'}^{\sigma\sigma}(\kx)
=\frac{1}{A_{BZ}}\int_{k_i(\kx)}^{k_f(\kx)}
\G_{\alpha\alpha'}^{\sigma\sigma}(\vec{k})e^{-i\kz (\zl-\zlp)}d\kz
\label{eqn:ift}
\end{equation}
for $\vec{k}=(\kx\hat{x},\kz\hat{z})$ in the Brillouin zone (BZ), $A_{BZ}$
the area of the BZ and $k_i(\kx)$ and $k_f(\kx)$ showing the initial and
final values of $\kz$ at each $\kx$. $\G_{\alpha\alpha'}^{\sigma\sigma}(\vec{k})$ has a
second order pole at $k=\sqrt{\kx^2+\kz^2}=0$ which is responsible for the
logarithmic long range behavior of LGF in real space.  The LGF in
reciprocal space is
\begin{equation*}
\G^{\sigma\sigma}_{\alpha\alpha'}(\kv)
=\frac{\hat{\G}_{\alpha\alpha'}^{\sigma\sigma}(\khat)}{\kx^2+\kz^2}f_c(k)
+\check{\G}_{\alpha\alpha'}^{\sigma\sigma}(\kv)
\end{equation*}
where $\hat{\G}^{\sigma\sigma}$ is the $\khat$ direction-dependent elastic Green's
function and $f_c(k)$ is a cutoff function that vanishes smoothly at the
edges of the BZ. In general anisotropic cases, $\hat{\G}^{\sigma\sigma}(\khat)$ is represented
by a Fourier series expansion as $\hat{\G}_{\alpha\alpha'}^{\sigma\sigma}(\khat)=\displaystyle
\sum_{n=0}^{N_{\text{max}}}\hat{\G}_{\alpha\alpha'}^{\sigma\sigma,n}e^{in\phi_k}$ where $\phi_k$ is the angle
of $\kv$ relative to an arbitrary in-plane direction and the truncation
$N_{\text{max}}$ is sufficiently large\cite{ref:lgf}.  The integrand in
\eqn{ift} is not singular for $\kx \neq 0$ however the $k^2$ pole in
$\G^{\sigma\sigma}(\vec{k})$ results in a pole of order $\left|\kx\right|$
in $\G_{\alpha l,\alpha'l'}^{\sigma\sigma}(\kx)$. To treat the small $\kx$
behavior analytically, we integrate \eqn{ift} as four terms
\begin{eqnarray}
\label{eqn:separation}
\int_{k_i(\kx)}^{k_f(\kx)}\G_{\alpha\alpha'}^{\sigma\sigma}(\vec{k})e^{-i\kz (\zl-\zlp)}d\kz & = &\int_{k_i(\kx)}^{k_f(\kx)}\G_{\alpha\alpha'}^{\sigma\sigma} (\vec{k})(e^{-i\kz (\zl-\zlp)}-1)d\kz \\ \nonumber
 &+ &\int_{k_i(\kx)}^{k_f(\kx)}\G_{\alpha\alpha'}^{\sigma\sigma}(\vec{k})-\frac{\hat{\G}_{\alpha\alpha'}^{\sigma\sigma,0}}{\kx^2+\kz^2}f_c(\kx,\kz)d\kz\\ \nonumber
&+& \int_{k_i(\kx)}^{k_f(\kx)}\frac{\hat{\G}_{\alpha\alpha'}^{\sigma\sigma,0}}{\kx^2+\kz^2}(f_c(\kx,\kz)-1)d\kz\\ \nonumber
&+& \int_{k_i(\kx)}^{k_f(\kx)}\frac{\hat{\G}_{\alpha\alpha'}^{\sigma\sigma,0}}{\kx^2+\kz^2}d\kz \\ \nonumber
\end{eqnarray}
where $\hat{\G}_{\alpha\alpha'}^{\sigma\sigma,0}$ is the $n=0$ coefficient in Fourier expansion of
$\hat{\G}_{\alpha\alpha'}^{\sigma\sigma}(\khat)$. The first three terms in
\eqn{separation} are evaluated numerically while the last integral is 
\begin{equation}
\label{eqn:pole}
\int_{k_i(\kx)}^{k_f(\kx)}\frac{\hat{\G}_{\alpha\alpha'}^{\sigma\sigma,0}}{\kx^2+\kz^2}d\kz=\frac{\pi \hat{\G}_{\alpha\alpha'}^{\sigma\sigma,0}}{|\kx|}+\hat{\G}_{\alpha\alpha'}^{\sigma\sigma,0}\left(\frac{\arctan(k_{f}(\kx)/\kx)-\arctan(k_i(\kx)/\kx)}{\kx}-\frac{\pi}{|\kx|}\right)
\end{equation}
where $\frac{\pi\hat{\G}_{\alpha\alpha'}^{\sigma\sigma,0}}{|\kx|}$ is the pole and the remaining terms are
added to the numerically evaluated part.  We add an elastic correction term
to $\G_{\alpha\alpha'}^{\sigma\sigma}(\kx)$, due to the interface obtained from Eqn. (59) in
\cite{ref:tewary}. Combining \eqn{pole}, \eqn{separation}, and
\eqn{elas_sep} produces
\begin{equation}
\label{eqn:gbb}
\G_{\alpha l,\alpha l'}^{\sigma\sigma'}(\kx)=\frac{\hat{\G}_{\alpha\alpha'}^{\sigma\sigma'}}{\left|\kx\right|}+\check{\G}_{\alpha l,\alpha l'}^{\sigma\sigma'}(\kx).
\end{equation}

\eqn{DG} has unknown blocks $\G^{II}(\kx)$, $\G^{I\sigma}(\kx)$.  Direct
substitution of the block partitions gives
\begin{eqnarray}
\label{eqn:gib}
\G^{I\sigma}(\kx)&=&-\Dinv\displaystyle\sum_{\sigma'=\pm}\tD^{I\sigma'}(\kx)\G^{\sigma'\sigma}(\kx) \\
\label{eqn:gii}
\Gii(\kx)&=&\Dinv+\displaystyle\sum_{\sigma'\sigma=\pm}\Dinv\tD^{I\sigma}(\kx)\G^{\sigma\sigma'}(\kx)\tD^{\sigma' I}(\kx)\Dinv.
\end{eqnarray} 
Note that by choosing the appropriate set of independent equations we
manage to avoid the calculation of the infinite dimensional
$\tD^{\sigma\sigma'}(\kx)$.  The finite range of $\tD^{I \sigma}(\kx)$
means that only a finite subset of atoms in each semi-infinite $\pm$ region
are considered for $\G^{\sigma\sigma'}(\kx)$.  To treat the poles in
$\Gii(\kx)$ and $\G^{I\sigma}(\kx)$ analytically, we use a  $\kx$
expansion of $\tD(\kx)=\hat{\tD}+\check{\tD}(\kx)$ derived from
\eqn{pft} where $\check{\tD}(\kx)=\tD^1\kx+O(\kx^2)$ . Therefore, for small $\kx$
\begin{eqnarray}
(\tD(\kx))^{-1}&=&\left[\hat{\tD}+\check{\tD}(\kx)\right]^{-1}\\ \nonumber
     &=&{\hat{\tD}}^{-1}\left[\mathbf{I}+\check{\tD}(\kx){\hat{\tD}}^{-1}\right]^{-1} \\ \nonumber
     &=& {\hat{\tD}}^{-1}-\kx{\hat{\tD}}^{-1}\tD^1{\hat{\tD}}^{-1}+O(\kx^2). \nonumber
\end{eqnarray}
 Using the small $\kx$
expansions for the bulk Green's functions with \eqn{gib} and \eqn{gii}
gives 
\begin{equation}
\label{eqn:gii_sep}
\Gib(\kx)=\frac{1}{\left|\kx\right|}\hat{\G}^{I\sigma}+\check{\G}^{I\sigma}(\kx) \ \text{and} \ 
\Gii(\kx)=\frac{1}{\left|\kx\right|}\hat{\G}^{II}+\check{\G}^{II}(\kx)
\end{equation}
where
\begin{equation*}
\hat{\G}^{I\sigma}=
-(\hat{\tD}^{II})^{-1}\sum_{\sigma'=\pm}\hat{\tD}^{I\sigma'}\hat{\G}^{\sigma'\sigma}
\text{ and } 
\hat{\G}^{II}= (\hat{\tD}^{II})^{-1}
+\sum_{\sigma,\sigma'=\pm}(\hat{\tD}^{II})^{-1}\hat{\tD}^{I\sigma}
\hat{\G}^{\sigma\sigma'}
\hat{\tD}^{\sigma'I}(\hat{\tD}^{II})^{-1}
\end{equation*}
are the constant coefficients of the pole and $\check{\G}^{I\sigma}(\kx)$
and $\check{\G}^{II}(\kx)$ include the remaining nonsingular terms.
$\check{\G}^{II}(\kx)$ and $\check{\G}^{I\sigma}(\kx)$ have a cusp approaching $\kx=0$ and the value at $\kx=0$ is
\begin{eqnarray}
\check{\G}^{I\sigma}(0)&=&-(\hat{\tD}^{II})^{-1}\sum_{\sigma^\prime=\pm}\hat{\tD}^{I\sigma^\prime}\check{\G}^{\sigma^\prime\sigma}(0) \\
\check{\G}^{II}(0)&=&(\hat{\tD}^{II})^{-1}+\sum_{\sigma,\sigma'=\pm}(\hat{\tD}^{II})^{-1}\hat{\tD}^{I\sigma}
\check{\G}^{\sigma\sigma'}(0)
\hat{\tD}^{\sigma'I}(\hat{\tD}^{II})^{-1}
\end{eqnarray}
where $\check{\G}^{\sigma\sigma^\prime}(0)$ is calculated in Appendix~\ref{sec:kx0}.
To ensure a smooth transition between interface and bulk regions, we compare the pole terms and the cusps for atom indices at the boundary between the regions (i.e $l_+$ and $l_-$). Labeling $l_\sigma$ as $\sigma=\pm$ or (I) does not
change the material response.  Specifically we should have 
\begin{eqnarray}
\label{eqn:test1}
 \hat{\G}_{\alpha l_\sigma,\alpha^\prime l_\sigma}^{II}=\hat{\G}_{\alpha\alpha^\prime}^{\sigma\sigma}&,&
\check{\G}_{\alpha l_\sigma,\alpha^\prime l_\sigma}^{II}(0)=\check{\G}_{\alpha l_\sigma,\alpha^\prime l_\sigma}^{\sigma\sigma}(0)
\end{eqnarray}
\eqn{test1}  determines the finite size effect in the interface.   Note that once the bulk force constant matrix is known, identifying atoms in the interface region does not require additional computation effort.

Evaluating the Green's function in real space between to atoms
$(\xl+na_0,\zl)$ and $(\xlp,\zlp)$ requires a partial inverse Fourier
transform over \eqn{gii_sep},
\begin{equation}
 \label{eqn:final_ift_ib}
 \GR_{\alpha\alpha'}^{I\sigma}(\xl-\xlp+na_0;\zl,\zlp)
=\int_{-k_{\text{max}}}^{k_{\text{max}}}
\G_{\alpha l,\alpha' l'}^{I\sigma}(\kx)e^{-i\kx (\xl - \xlp + na_0)}d\kx
\end{equation}
and
\begin{equation}
 \label{eqn:final_ift_ii}
 \GR_{\alpha\alpha'}^{II}(\xl-\xlp+na_0;\zl,\zlp)
=\int_{-k_{\text{max}}}^{k_{\text{max}}}
\G_{\alpha l,\alpha' l'}^{II}(\kx)e^{-i\kx (\xl-\xlp+na_0)}d\kx.
\end{equation}
The $\hat{\G}$ term in \eqn{gii_sep} is treated analytically via
\begin{equation*}
\int_{-\infty}^{\infty}\frac{1}{\left|\kx\right|}e^{-i\kx
x}d\kx=-2\ln\left|x\right|.
\end{equation*}
Therefore
\begin{equation}
\label{eqn:pole-ift}
 \int_{-k_{\text{max}}}^{k_{\text{max}}}{\frac{1}{|k_{x}|}e^{-ik_{x}x}}d\kx=-2\ln|x|+2\text{Ci}( k_{\text{max}}x).
\end{equation}
Note that $\lim \limits_{x \to 0}
-2\ln|x|+2\text{Ci}=2\gamma+2\ln(k_{\text{max}})$ where $\gamma\approx
0.577215$ is the Euler constant.  The partial inverse Fourier transform for
$\check{\G}$ terms are evaluated numerically over a discrete $\kx$ mesh of
size $N_{\kx}$
\begin{equation}
\label{eqn:pift1}
\check{\G}_{\alpha\alpha'}^{I\sigma}(\xl-\xlp+na_0;\zl,\zlp)=
\frac{1}{N_{\kx}}\displaystyle \sum_{\kx}\check{\G}_{\alpha l,\alpha'
l'}^{I\sigma}(\kx)e^{-i\kx(\xl-\xlp+na_0)}
\end{equation}
and
\begin{equation}
\label{eqn:pift2}
\check{\G}_{\alpha\alpha'}^{II}(\xl-\xlp+na_0;\zl,\zlp)=
\frac{1}{N_{\kx}}\displaystyle \sum_{\kx}\check{\G}_{\alpha l,\alpha'
l'}^{II}(\kx)e^{-i\kx(\xl-\xlp+na_0)}.
\end{equation}
\tab{summary} summarizes the method.
\begin{table}[h]

\caption{Summary of the procedure for ILGF computation. Regions (+, $-$,
and I) are defined in \fig{interface}b. 
$\GR^{\--+}(x;z,z')$ is the elastic Green's function for
a bicrystal computed by Tewary \et\cite{ref:tewary}.
$\G^{\sigma\sigma}(\kv)$ is the  LGF in bulk $\sigma=\pm$.  FT prefactors required to maintain the consistency between elastic bicrystal GF and bulk LGF solutions are also listed.}

\begin{enumerate}
\label{tab:summary}
\item
Compute $\D_{\alpha\alpha'}(\xl-\xlp+na_0;\zl,\zlp)$ directly. Divide the
geometry into $-,\text{I},+$ regions.
\item
$\tD_{\alpha l,\alpha' l'}^{I\sigma}(\kx)
=\displaystyle\sum_{n=-\infty}^{\infty}e^{i\kx
(\xl-\xlp+na_0)}\D_{\alpha\alpha'}^{I\sigma}(\xl-\xlp+na_0;\zl,\zlp)$, $a_0=$ periodicity in $x$ direction and $\sigma=\pm,\text{I}$.~\eqn{pft}
\item
$\G_{\alpha l,\alpha' l'}^{-+}(\kx)=\frac{b_1}{2\pi
\left|\ttwo\right|}\displaystyle
\int_{-\infty}^{\infty}\GR_{\alpha\alpha'}^{\--+}(x;z_l,z_{l'})
e^{i\kx x}dx$, $b_1=\frac{2\pi}{a_0}$, $b_1 b_2=A_{BZ}$.~\eqn{gft}
\item
$\G_{\alpha l,\alpha' l'}^{\sigma \sigma}(\kx)=\frac{1}{b_2}\displaystyle
\int_{k_i(\kx)}^{k_f(\kx)}\G_{\alpha\alpha'}^{\sigma \sigma}(\kx;\kz)
e^{-i\kz (z_l-z_{l'}) }d\kz$.~\eqn{ift}
\item
$\G_{\alpha l,\alpha' l'}^{\sigma \sigma'}(\kx)=\frac{\hat{\G}_{\alpha
\alpha'}^{\sigma \sigma'}}{\left|\kx\right|}+\check{\G}_{\alpha
l,\alpha' l'}^{\sigma \sigma'}(\kx)$.~\eqn{gbb}
\item
$\tD_{\alpha l,\beta n}(\kx)\G_{\beta n,\alpha' l'}(\kx)=\delta_{\alpha
\alpha'}\delta_{l l'}$ $\longrightarrow$ $\G_{\alpha
l,\alpha' l'}^{\text{I}\sigma}(\kx)=\frac{\hat{\G}_{\alpha
\alpha'}^{\text{I}\sigma}}{\left|\kx\right|}+\check{\G}_{\alpha
l,\alpha' l'}^{\text{I}\sigma}(\kx)$,
($\sigma=\pm,\text{I}$).~\eqn{gib}-(\ref{eqn:gii_sep})
\item
$\G_{\alpha l,\alpha' l'}^{I\sigma}(x=\xl-\xlp+na_0;z_l,z_{l'})$\\
$=\frac{\hat{\G}_{\alpha
\alpha'}^{\text{I}\sigma}}{b_1}\left(-2\ln|x|+2\text{Ci}(\frac{b_1}{2}
x)\right)+\frac{1}{N_{\kx}}\!\displaystyle
\sum_{m=1}^{N_{\kx}-1}\check{\G}_{\alpha l,\alpha'
l'}^{I\sigma}(\frac{mb_1}{N_{\kx}})e^{-i\frac{mb_1}{N_{\kx}}x}$.~\eqn{pole-ift}-(\ref{eqn:pift2})
\end{enumerate}
\end{table}

\section{Application: Lattice Green's function for T\lowercase{i} $(10\bar{1}2)$ twin boundary}
\label{sec:application}
We use the method to compute the ILGF for a Ti lattice containing
$(10\bar{1}2)$ twin boundary.  The geometry of this boundary is shown in
\fig{interface}c.  The $\underline{F}^{\lambda,\mu}$ and
$\underline{M}^{\lambda,\mu}$ matrices are
\begin{equation*}
\underline{F}^{\lambda,\mu}=\mathbf{I} , \ 
\underline{M}^\lambda=
\left[\begin{array}{c c c}
 2 & 0 & 0\\
 1 & 1 & 0 \\
 1 & 0 & 1
\end{array}\right],
\text{ and }
\underline{M}^\mu=
\left[\begin{array}{ccc}
 2 & 0 & 0\\
 1 & 1 & 0\\
-1 & 0 & 1
\end{array}\right].
\end{equation*}
The twin boundary is defined by $\tone=\sqrt{3a^2+c^2}\hat{x}$ and
$\ttwo=a\hat{y}$ where $a$ and $c$ are the hcp lattice constants in
Ti. Lattice $\mu$ is the reflection of $\lambda$ about the twin boundary
plane.  The force-constant matrices $\underline{D}(\R)$ are computed using
\textsc{lammps} package\cite{ref:LAMMPS} with a Ti MEAM
potential with the maximum cut off distance of 5.5\AA\cite{ref:meam}. The partial FT in \eqn{pft} is done by a uniform
discrete mesh of 40 $\kx$ points over $(-\pi/a_0,\pi/a_0)$ where $a_0$ is
the periodicity of the geometry in $x$ direction and equal to
$\left|\tone\right|$ in this case. The same $\kx$ values must be used in
$(+)$,$(-)$ and (I) regions. Limits of $k_z$ in \eqn{ift} are then chosen
so that the equivalent of $A_{BZ}$ is covered in both $(+)$ and $(-)$. The
first three integrals in \eqn{separation} are evaluated numerically over a
uniform $\kz$ mesh of 160 points at each $\kx$. For $\left|\kx\right|< 0.1
\pi/a_0$, the density of $\kz$ mesh is doubled to insure the convergence
around the discontinuity at $\Gamma$-point\cite{ref:lgf,ref:effic}.
\begin{figure} \begin{center}
   \includegraphics[width=5 in]{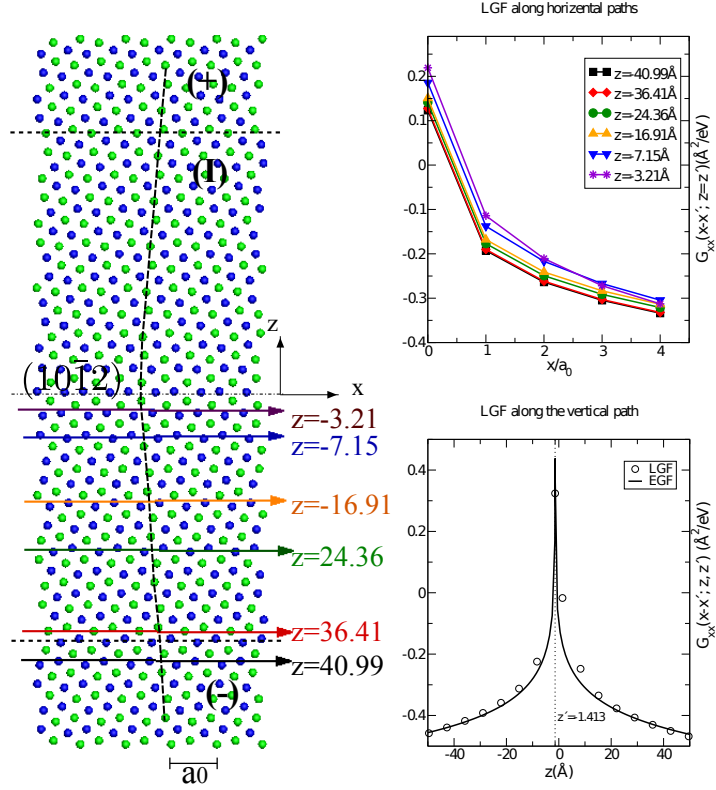}
 \end{center}

\caption{$\left[1\bar{2}10\right]$ projection of the Ti supercell
containing a $\left(10\bar{1}2\right)$ twin boundary. The supercell is
divided into bulk ($+/-$) and interface (I) regions. $y$ axis is pointing into the
plane.  Variation of the $G_{xx}$ component of the lattice Green's function
is plotted along six horizontal and one vertical paths. The reference atom
$\left(x',z'\right)$ is the first atom in horizontal paths and
the atom right below the interface in the vertical path.  Bulk behavior
along the $z=z'$ paths is recovered away from the interface. The long
range behavior of the LGF matches the EGF along the vertical path, while
deviating for small $z-z'$.  }
\label{fig:TB}
\end{figure}
\indent \fig{TB} shows the supercell with bulk ($+/-$) and interface (I) divisions
and the paths along which LGF is evaluated for testing purposes.
$G_{xx}(x-\xlp;\zl,\zlp)$ is plotted along a vertical and six horizontal
paths  in the supercell where the reference atom  $l^\prime$ is
the first atom ($\xlp=0$) in the horizontal paths and the atom right below
the interface in the vertical path.  Bulk response along $\zl=\zlp$ paths is
gradually recovered as the paths get farther from the interface and closer
to the ($-$) region. In addition, it is worth noting that paths 1 and 2 are
located in bulk and interface regions respectively. Therefore, the LGF is
obtained from the bulk lattice Green's function along path 1 and from the
ILGF method along path 2. The good agreement between the response of these
two paths verifies the smooth transition between the bulk-interface
divisions. $G_{xx}(x-x';z,z^\prime)$ as a function of $z$ is also plotted
for atoms along the vertical line shown on the supercell in
\fig{TB}. The reference atom is located on the vertical line at
$\xlp=x',\zlp=z'=-1.413\text{\AA}$ which is right below the interface. The long range
behavior of the ILGF matches the EGF.

We apply the computed ILGF to simulate the interaction of a
$[\bar12\bar10]$ screw dislocation with the Ti$(10\bar12)$ twin boundary by
flexible boundary conditions \cite{ref:Sinclair,ref:woodward} with a Ti MEAM potential\cite{ref:meam}.  Periodic
boundary conditions are applied along the dislocation line.  Flexible
boundary conditions relax atoms surrounding the dislocation core region
with the lattice Green's function as if they are embedded in an infinite
medium. Conjugate-gradient method relaxes the atoms around the dislocation core (region 1). This process generates forces on atoms of the intermediate
region (region 2). ILGF relaxes the forces on region 2 and updates the
positions of the outermost atoms (region 3), originally obtained from the
elastic displacement field of the screw dislocation.  To verify the
results, we also modeled the same dislocation/interface geometry with fixed
boundary conditions using supercell radii of 12--50b; b is the magnitude of
the Burgers vector equal to $\left|\ttwo\right|$.  Outer layers of atoms in
a region of width 3b are frozen to elastic displacement field of the
screw dislocation and the inner atoms are relaxed through the
conjugate-gradient method using Ti MEAM.  Large supercells are required to
minimize the effect of free surfaces created by the fixed boundaries.

\begin{figure} \begin{center}
   \includegraphics[angle=270,width=\textwidth]{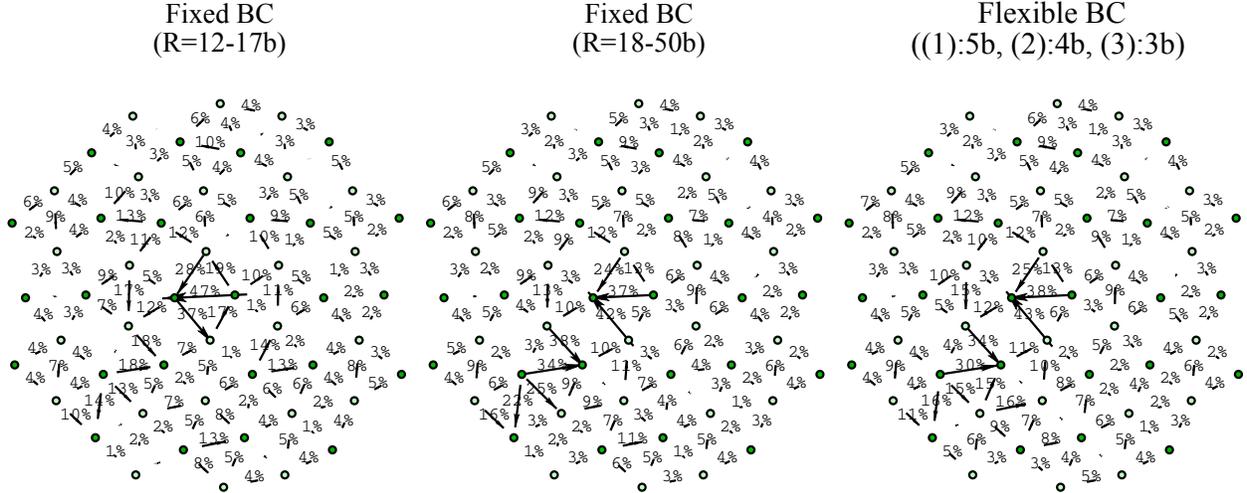}
 \end{center}

\caption{Differential displacement maps of a screw dislocation core in Ti
$\left(10\bar{1}2\right)$ twin boundary computed by fixed and flexible
boundary conditions. Fixed boundary conditions cause a supercell size effect which is
evident from different core structures for radius $R$ smaller or larger
than 17b (1312 atoms relaxed). Flexible boundary conditions give the same core structure as
the large fixed boundary conditions supercell with significantly less atoms required to
relax by energy minimization (i.e  73 atoms in region (1) and 652 atoms total).
}
\label{fig:screw}
\end{figure}

\fig{screw} shows the differential displacement maps\cite{ref:DD} of the
screw dislocation core structure in the Ti $(10\bar{1}2)$ twin boundary
obtained by fixed and flexible boundary. Fixed boundary conditions result in a finite size
effect that is removed with flexible boundary conditions, or with significantly larger
calculations.  For supercell radii $R\leq 17\text{b}$ ($R$=17b corresponds to 1312
atoms relaxed), the dislocation center is trapped in the interface while
for $R$ between 18 and 50b---corresponding to 1474 and 11364 atoms
respectively---the dislocation center moves out of the interface towards
the bottom lattice.  This is possible due to the broken mirror symmetry at
the twin boundary for this MEAM potential.  The flexible boundary conditions supercell has
$R$=12b and 652 ((1):73, (2):219, (3): 360) atoms. 
The core structure from
flexible boundary conditions is in good agreement with large fixed boundary conditions results-- hence  the correct structure can be obtained using flexible boundary conditions with
significantly less atoms than with fixed boundary conditions.

\section{Conclusions}
\label{sec:conclusions}  
We developed an automated computational approach to calculate the lattice
Green's function of crystals containing planar interfaces for arbitrary
force constants and interface orientations. This method is more general
than the previous Dyson-equation approaches in the sense that it can
consider long range atomic interactions and reconstructions near the
interface. We computed the ILGF for a Ti $\left(10\bar{1}2\right)$ twin
boundary with a Ti MEAM potential and studied the screw dislocation/twin
boundary interaction using flexible boundary conditions.  Our results show
that the ILGF flexible boundary conditions method predicts the correct
dislocation core structure.  Moreover, the energy minimization stage of the
flexible boundary conditions involves significantly less atoms than what is required by
fixed boundary conditions methods. This highlights the applicability of flexible boundary conditions methods
to modeling defect/interface interactions by DFT.
\section{Acknowledgments}
This work is supported by NSF/CMMI grant 0846624.
\appendix
\section{Evaluation of $\mathbf{\check{\G}_{\alpha l,\alpha^\prime l^\prime}^{\sigma\sigma^\prime}(\kx=0)}$ }
\label{sec:kx0}
\subsection{$\sigma=\sigma^\prime$}
$\check{\G}_{\alpha l,\alpha^\prime l^\prime}^{\sigma\sigma}(\kx=0)$ is obtained by taking the limit of \eqn{separation} and \eqn{pole} as $\kx\rightarrow 0$:
\begin{eqnarray}
\label{eqn:kx1}
\check{\G}_{\alpha l,\alpha^\prime l^\prime}^{\sigma\sigma}(\kx=0) & = &\int_{k_i(0)}^{k_f(0)}\G^{\sigma\sigma} (\kz\hat{z})(e^{-i\kz (\zl-\zlp)}-1)d\kz \\ 
\label{eqn:kx2}
 &+ &\int_{k_i(0)}^{k_f(0)}\G^{\sigma\sigma}(\kz\hat{z})-\frac{\hat{\G}^{\sigma\sigma}}{\kz^2}f_c(\kz)d\kz\\ 
\label{eqn:kx3}
&+& \int_{k_i(0)}^{k_f(0)}\frac{\hat{\G}^{\sigma\sigma}}{\kz^2}(f_c(\kz)-1)d\kz\\ 
\label{eqn:kx4}
&+& \displaystyle\lim_{\kx\rightarrow 0}\hat{\G}^{\sigma\sigma}\left(\frac{\arctan(k_{f}(\kx)/\kx)-\arctan(k_i(\kx)/\kx)}{\kx}-\frac{\pi}{|\kx|}\right).
\end{eqnarray}
Note that since $\kv=\kz\hat{z}$, $\hat{\G}^{\sigma\sigma}(\khat)$ is evaluated along a constant $\khat$-direction and therefore is a constant. The cut off function is

$f_c(\kz)=\Bigg\{$\begin{tabular}{lll}
1&    &  $0<\left|\kz\right|<0.5 \kz^\text{max}$ \\
$12 (1-\left|\kz\right|)^2-16(1-\left|\kz\right|)^3 $& & $0.5 \kz^\text{max}< \left|\kz\right|< \kz^\text{max} $ 

\end{tabular}
\\

\noindent where $\kz^\text{max}\le \text{Min}(\left|k_i(0)\right|,k_f(0))$
to insure that $f_c(\kz)=0$ at the Brillouin zone boundary. We isolate the
$\kz=0$ point by dividing the integration path in \eqn{kx1}, \eqn{kx2} and
\eqn{kx3} into three intervals
\[
\left[k_i(0),k_f(0)\right]=\left[k_i(0),-\epsilon/2\right)\cup\left[-\epsilon/2,\epsilon/2\right]\cup\left(\epsilon/2,k_f(0)\right]
\]
where $\epsilon$ is sufficiently small.  The first and third intervals
do not contain the $\Gamma$-point and therefore their corresponding
integrals are evaluated numerically without special treatments. To evaluate
the integrals in \eqn{kx1} and \eqn{kx2} over
$\left[-\epsilon/2,\epsilon/2\right]$, we use the small $\kz$ leading order
terms of $\G^{\sigma\sigma}(\kv)$ \cite{ref:lgf} and the exponential term
\begin{multline*}
\int_{-\epsilon/2}^{\epsilon/2}\G^{\sigma\sigma} (\kz\hat{z})(e^{-i\kz
(\zl-\zlp)}-1)d\kz=\\
\int_{-\epsilon/2}^{\epsilon/2}\left(\frac{\hat{\G}_{\alpha\alpha'}^{\sigma\sigma}}{\kz^2}+\frac{i}{\left|\kz\right|}{\G}_{\alpha\alpha'}^{\sigma\sigma,i}\frac{\kz}{\left|\kz\right|}+\G_{\alpha\alpha'}^{\text{D}}(\kz)\right)\left(-i\kz
(\zl-\zlp)-\kz^2\frac{(\zl-\zlp)^2}{2}\right)d\kz= \\
\left({\G}_{\alpha\alpha'}^{\sigma\sigma,i}(\zl-\zlp)-\hat{\G}_{\alpha\alpha'}^{\sigma\sigma}\frac{(\zl-\zlp)^2}{2}\right)\epsilon
\end{multline*}
and
\[
\int_{-\epsilon/2}^{\epsilon/2}\G^{\sigma\sigma}(\kz\hat{z})-\frac{\hat{\G}^{\sigma\sigma}}{\kz^2}f_c(\kz)d\kz=
\int_{-\epsilon/2}^{\epsilon/2}\left(i\frac{{\G}_{\alpha\alpha'}^{\sigma\sigma,i}}{\kz}+\G_{\alpha\alpha'}^{\text{D}}(\kz)\right)d\kz=\G_{\alpha\alpha'}^{\text{D}}(0)\epsilon.
\]
$\hat{\G}_{\alpha\alpha'}^{\sigma\sigma}/\kz^2$ and
$\G_{\alpha\alpha'}^{\text{D}}(\kz)$ are the elastic and discontinuity
corrections and $i{\G}_{\alpha\alpha'}^{\sigma\sigma,i}/\kz$ appears only
in the case of a multiatom basis. $\hat{\G}_{\alpha\alpha'}^{\sigma\sigma}$
and $\G_{\alpha\alpha'}^{\sigma\sigma,i}$ are constants
here\cite{ref:lgf,ref:effic}. Also note that $f_c(\kz)=1$ over
$\left[-\epsilon/2,\epsilon/2\right]$; hence the integral in \eqn{kx3}
equals zero over this interval.

  Taking $\epsilon$ to be $\frac{k_f(0)-k_i(0)}{N_{\text{div}}}$ where $N_{\text{div}}$ is the number of divisions in the discrete $\kz$ mesh we have
\begin{eqnarray}
\nonumber
\check{\G}_{\alpha l,\alpha^\prime l^\prime}^{\sigma\sigma}(0)&=&\frac{k_f(0)-k_i(0)}{N_{\text{div}}}\Bigg[\sum_{\kz\ne 0}\left( \G_{\alpha\alpha'}^{\sigma\sigma}(\kz\hat{z})e^{-i\kz (\zl-\zlp)}-\frac{\hat{\G}^{\sigma\sigma}}{\kz^2}\right)\\ \nonumber
& +&\G_{\alpha\alpha'}^{\sigma\sigma,i}(\zl-\zlp)-\frac{\hat{\G}_{\alpha\alpha'}^{\sigma\sigma}(\zl-\zlp)^2}{2}+\G_{\alpha\alpha'}^{\text{D}}(0)\Bigg]\\ \nonumber
&+&\hat{\G}_{\alpha\alpha'}^{\sigma\sigma}\left(\frac{1}{k_i(0)}-\frac{1}{k_f(0)}\right).
\end{eqnarray}
The first summation

\begin{tabular}{l}
\\
$\displaystyle\frac{k_f(0)-k_i(0)}{N_{\text{div}}}\sum_{\kz\ne 0}\left( \G_{\alpha\alpha'}^{\sigma\sigma}(\kz\hat{z})e^{-i\kz (\zl-\zlp)}-\frac{\hat{\G}^{\sigma\sigma}}{\kz^2}\right)=$\\
$\displaystyle\frac{k_f(0)-k_i(0)}{N_{\text{div}}}\sum_{\kz\ne 0}{\left( \G_{\alpha\alpha'}^{\sigma\sigma}(\kz\hat{z})(e^{-i\kz (\zl-\zlp)}-1)\right)+\left(\G^{\sigma\sigma}(\kz\hat{z})-\frac{\hat{\G}^{\sigma\sigma}}{\kz^2}f_c(\kz)\right)+\left(\frac{\hat{\G}^{\sigma\sigma}}{\kz^2}(f_c(\kz)-1)\right)}$\\
\\
\end{tabular}

\noindent is the numerical integration of all three integrals in \eqn{kx1}-(\ref{eqn:kx3}) over $\left[k_i(0),-\epsilon/2\right)\cup\left(\epsilon/2,k_f(0)\right]$. The last term $\hat{\G}_{\alpha\alpha'}^{\sigma\sigma}\left(\frac{1}{k_i(0)}-\frac{1}{k_f(0)}\right)$ is the evaluation of \eqn{kx4}.

\subsection{$\sigma\ne\sigma^\prime$}
$\check{\G}_{\alpha\alpha'}^{\sigma\sigma^\prime}(\kx=0)$ is obtained from the small $\kx$ expansion of \eqn{bipft} and removing the $\kx^{-1}$ term
\begin{equation}
\check{\G}_{\alpha\alpha'}^{\sigma\sigma^\prime}(0)=\frac{\pi}{\left|\ttwo\right|}\left(2a_{\alpha\alpha^\prime}^{qq^\prime}\left|\beta_{l,l^\prime}^{qq^\prime}\right|-b_{\alpha\alpha^\prime}^{qq^\prime}\gamma_{l,l^\prime}^{qq^\prime}\right).
\end{equation}


\newpage

\end{document}